\documentclass[12pt]{article}
\usepackage[a4paper,margin=1.25in,footskip=.5in]{geometry}
\usepackage{authblk}
\usepackage[latin9]{inputenc}
\usepackage{float}
\usepackage{units}
\usepackage{siunitx}
\usepackage[version=4]{mhchem}
\usepackage{hyperref}
\hypersetup{pdftitle={Li-Si battery paper},
	    	pdfauthor={Ata Mesgarnejad, Northeastern University},
	    	colorlinks,
	    	pdfcreator={pdflatex},
	    	unicode=false,
	    	pdftoolbar=false,
	    	pdfmenubar=true,
	   		pdffitwindow=true,
	   		pdfnewwindow=true,
	    	linkcolor=red,
	    	citecolor=red,
	    	filecolor=black,
	    	urlcolor=blue,
	    	}
\usepackage{mathrsfs}
\usepackage{amsthm}
\usepackage{amsmath}
\usepackage{amssymb}
\usepackage{graphicx}
\usepackage{pbox}
\usepackage{esint}
\usepackage{nomencl}
\usepackage{diagbox,ragged2e}
\usepackage{algpseudocode}
\makeatletter
\usepackage[normalem]{ulem}

\floatstyle{ruled}

\newfloat{algorithm}{tbp}{loa}
\floatname{algorithm}{Algorithm}
\theoremstyle{plain}

\theoremstyle{definition}

\theoremstyle{remark}

\usepackage{booktabs}
\usepackage{amsxtra}
\usepackage{mathrsfs}
\usepackage{amsthm}
\usepackage{pdfsync}

\usepackage{xargs}                      
\usepackage[pdftex,dvipsnames]{xcolor}
\newcommandx{\AM}[1]{\textcolor{NavyBlue}{#1}}

\newcommand{\F}{\mathscr{F}}

\newcommand{\thth}{\theta\theta}
\newcommand{\sy}{\sigma_y}

\newcommand{\ie}{\emph{i.e., }}

\newcommand{\El}{\mathrm{E}}
\newcommand{\Fig}{Figure}

\usepackage{zref-user}
\usepackage{zref-xr}
\zxrsetup{tozreflabel}
\zexternaldocument*{Mesgarnejad-Karma-Li-Si-Supp}
\newcommand{\myzref}[1]{\textcolor{red}{\zref{#1}}}

\title{Vulnerable Window of Yield Strength for Swelling-Driven Fracture of Phase-Transforming Battery Materials}

\author[1]{Atoallah Mesgarnejad}
\author[1]{Alain Karma\footnote{a.karma@northeastern.edu}}

\affil[1]{Center for Inter-disciplinary Research on Complex Systems, Department of Physics, Northeastern University, Boston, MA. 02115, U.S.A.}

\begin{document}
 \maketitle
\begin{abstract}
Despite numerous experimental and theoretical investigations of the mechanical behavior of high-capacity Si and Ge Li-ion battery anodes, our basic understanding of swelling-driven fracture in these materials remains limited. Existing theoretical studies have provided insights into elasto-plastic deformations caused by large volume change phase transformations, but have not modeled fracture explicitly beyond Griffith's criterion. Here we use a multi-physics phase-field approach to model self-consistently anisotropic phase transformation, elasto-plastic deformation, and crack initiation and propagation during lithiation of Si nanopillars. Our computational results reveal that fracture occurs within a ``vulnerable window'' inside the two-dimensional parameter space of yield strength and fracture energy and highlight the importance of taking into account the surface localization of plastic deformation to accurately predict the magnitude of tensile stresses at the onset of fracture. They further demonstrate how the increased robustness of hollow nanopillars can be understood as a direct effect of anode geometry on the size of this vulnerable window. Those insights provide an improved theoretical basis for designing next-generation mechanically stable phase-transforming battery materials undergoing large volume changes.
\end{abstract}

\section*{Introduction}

Increasing demand for portable energy storage has motivated a large research activity focused on high-capacity Li-ion battery anodes. 
Current carbon-based anodes have limited theoretical capacity (\SI{372}{\milli\ampere\hour\per\gram} for \ce{Li6C}~\cite{Kasavajjula:2007}). 
Silicon and germanium have an order of magnitude larger theoretical capacity (\SI{3579}{\milli\ampere\hour\per\gram} for \ce{Li15Si4}, \SI{4200}{\milli\ampere\hour\per\gram} for \ce{Li22Si5}~\cite{Kasavajjula:2007}, \SI{1384}{\milli\ampere\hour\per\gram} for \ce{Li15Si4}~\cite{Liu:2014}) but are prone to fracture due to the high, approximately $300\%$, volume expansion during lithiation~\cite{Liu:2011a,Liu:2011b}, which limits their use. Different designs have been explored to overcome this limitation including silicon nanopillars~\cite{Liu:2012,Lee:2012}, thin films~\cite{Graetz:2004,Sethuraman:2010,Soni:2011}, open nano-porous crystalline Si structures with ultra-high interfacial area produced by dealloying of Si-based alloys~\cite{Zhao:2012a,Wada:2014}, combinations of these~\cite{Ge:2012,Zhang:2017}, as well as composite designs that embed silicon particles inside a more mechanically stable matrix ~\cite{Zhang:2006a,Guo:2010,Yue:2012,Terranova:2014,Tocoglu:2016}.

Basic studies of the lithiation process have shown that crystalline silicon (c-Si) transforms to an amorphous lithiated alloy (a-Li$_x$Si)~\cite{Lee:2011,Liu:2012a,Liu:2014}. The kinetics of this large volume change phase transformation is understood to be both interface-reaction limited~\cite{Liu:2012a} and highly anisotropic, reflecting the two key observations that the velocity of the c-Si/a-Li$_x$Si interface remains approximately constant during lithiation, and that this velocity depends strongly on crystallographic orientation~\cite{Lee:2011}. 
Numerical studies of elasto-plastic deformations of c-Si particles (nanowires, nano-/micro-pillars, etc) have demonstrated that the resulting anisotropic swelling can produce both large shape changes of the particles and, as a non-trivial effect of compressive yielding, tensile stresses on their outer surface that can potentially drive fracture~\cite{Liu:2011a,Yang:2012a,Yang:2014,An:2015}. Those insights have already proven useful to test new anode designs to mitigate fracture~\cite{An:2015}. However, our ability to predict when and how fracture occurs in different large volume change materials (e.g. Si versus Ge) and different anode geometries (e.g. solid versus hollow nanopillars~\cite{Zhang:2017}) is still limited. To date, the onset of fracture has been predicted using analytical solutions for stresses obtained in idealized geometries, assuming purely plastic deformation and isotropic swelling, and by applying a Griffith criterion to predict fracture onset for a flaw size comparable to the particle dimension~\cite{Zhao:2012,Zhao:2012b}. However, crack initiation and propagation in the setting of large elasto-plastic deformation and anisotropic swelling in phase-transforming materials remain largely unexplored.  

Here, we use a multi-physics phase-field approach to simulate both anisotropic swelling and fracture of solid and hollow c-Si nanopillars within a unified theoretical framework and derive from our simulations an understanding of when and how fracture occurs as a function of key materials parameters, including yield strength and fracture energy, and geometric parameters such as nanopillar radius and slenderness.  
Phase transformation is modeled using a phase-field $\psi$ that distinguishes the c-Si and a-Li$_x$Si phases and is evolved dynamically to describe the interface-reaction-limited anisotropic motion of the c-Si/a-Li$_x$Si interface. Fracture, in turn, is modeled using the well-established variational approach that couples elasticity to a phase-field $\phi$, which distinguishes pristine and broken regions of the material~\cite{Karma:2001a,Bourdin:2008a}. To realistically model large volume changes, this variational approach is implemented using a large deformation formulation of elasto-plasticity combining neo-Hookean nonlinear elasticity and $J_2$ plasticity to quasi-statically evolve $\phi$ together with the material displacement field and the plastic deformation gradient tensor. 
The phase-field approach offers several advantages in the present context. It provides a self-consistent formulation to model simultaneously anisotropic swelling, large elasto-plastic deformation, and fracture. Furthermore, it can describe the evolution of phase boundaries and cracks of arbitrarily complex shapes, as demonstrated in applications to other phase transformations~\cite{karma2016atomistic} and fracture problems such as thermal shock fracture~\cite{Bourdin:2014a}, mixed mode fracture~\cite{Chen:2015b}, ductile fracture~\cite{Ambati:2015,Borden:2016}, and the simpler chemo-mechanical fracture of single-phase battery cathode particles~\cite{Miehe:2015,Zuo:2015,Klinsmann:2016,Mesgarnejad:2019}, also driven by volume expansion due to Li intercalation but only involving small elastic stresses and no phase change. 
In addition, in contrast to Griffith theory, the phase-field approach is able to describe crack initiation without pre-existing flaws. This property stems from the fact that $\phi$ varies smoothly in space on a length scale $\xi$, thereby enabling crack formation on the scale of the ``process zone'' where elastic energy is transformed into new fracture surfaces. Hence, directly relevant to the present study, the phase-field approach can quantitatively describe crack initiation from surface imperfections such as U- or V-shape notches by treating $\xi$ treated as a material-dependent parameter~\cite{tanne2018crack}. V-shape notches, in particular, bear close similarity to surface shape deformations of lithiated Si particles undergoing elasto-plastic deformation during anisotropic swelling~\cite{Liu:2011a,Yang:2012a,Yang:2014,An:2015}. 

To keep computations tractable, we perform 2D plane-strain simulations ($\partial_z\equiv 0$) on a cross-section of an unconstrained nanopillar ($\tau_{zz}=0$) lithiated from its surface (i.e. outer boundary for a solid nanopillar and both outer and inner boundaries for hollow nanopillars).
Furthermore, to dissect the contributions of multiple interacting physical effects (including compressive and tensile yielding, anisotropic swelling, localization of plastic deformation, and crack initiation and propagation), we carry out three different types of computations of increasing complexity. 
In a first step, we model stress evolution without fracture by assuming that swelling is isotropic and that the stress fields and plastic hardening parameter $\alpha$ only vary radially and are independent of the azimuthal angle $\theta$ as depicted schematically in \Fig~\ref{fig:figure1}a. 
This axisymmetric approximation reduces the 2D problem to a 1D radial problem. 
Stress evolution in a similar idealized geometry has been previously studied analytically by taking into account only plastic deformation~\cite{Zhao:2012,Zhao:2012b}. 
By taking into account here both elastic and plastic deformations, we demonstrate that tensile stresses generated on the particle surface by volume expansion reach a maximum value as a function of yield strength $\sy$. 
Even though the critical $\sy$ value corresponding to this maximum is outside the experimentally~\cite{Sethuraman:2010,Chon:2011,Pharr:2013} or theoretically estimated~\cite{Zhao:2011c} range $\sy\sim \SIrange{0.5}{2}{\giga\pascal}$ for Si, the existence of this critical yield strength provides a valuable theoretical framework to understand fracture behavior inside this lower estimated range of $\SIrange{0.5}{2}{\giga\pascal}$. For this reason, we investigate stress evolution over a wide range of $\sy$ that encompasses the entire vulnerable window for fracture.
In a second step, we carry out a similar computation, still without fracture, but for the full 2D problem without the axisymmetric approximation in which the stress fields and $\alpha$ can vary both radially and azimuthally inside the pillar cross section. 
This enables us to asses how anisotropic swelling modifies tensile stresses on the pillar surface. We find that tensile stresses become amplified by localization of plastic deformation, but still exhibit a maximum as a function of increasing $\sy$. 
Those 1D and 2D computations demonstrate the existence of a vulnerable window of yield strength inside which pillars are prone to fracture. 
More crucially, for the relevant experimentally reported yield strengths ($\sy=\SIrange{0.5}{2}{\giga\pascal}$) of Si, the difference between the material and the most critical yield strength controls the magnification of  generated tensile stresses.
In a last step, we validate the existence of this window by repeating our 2D computations with fracture, showing that pillars fracture only over an intermediate range of yield strength.
We compare the results of full 2D simulations with estimates based on our numerically calculated stresses in the previous steps using the Griffith theory framework.
We then use experimental estimates of safe pillar radius (i.e. largest pillar radius without fracture) to quantitatively validate our findings by calculating our estimate of yield strength. 
\section*{Results and discussion}
\subsection*{Model}

We model the swelling-driven deformation of the material using the finite $J_2$ elasto-plasticity framework ~\cite{Simo:1988,Simo:1988a,Simo:2006} and account for the fracture of the material by coupling it to a phase-field fracture model \cite{Karma:2001a,Bourdin:2008a}.
Furthermore, we model the anisotropic motion of the c-Si/a-Li$_x$Si interface during lithiation using a non-conserved phase field $\psi$ where $\psi=0$ in the crystalline phase and $\psi=1$ in the amorphous phase.
The material properties are approximated using a linear role of mixture between the crystalline and amorphous phases. We define the deformation gradient tensor as $F_{ij}=\partial x_i/\partial X_j=1+\partial u_i/\partial X_j$ where $X_i$ are the undeformed coordinates and $x_i=X_i+u_i$ are the deformed coordinates of material points and $u$ is the displacement field. 
We use a multiplicative decomposition of the deformation gradient tensor such that
\begin{equation}\label{eq:mult-decomp}
 	F_{ij}=\sqrt{J_{\psi}}F^e_{ik}F^p_{kj}
\end{equation}
where $J_\psi=(1+\beta\psi)^2$ is the phase dependent volumetric expansion due to phase change with linear Vegard expansion coefficient $\beta$, $F^p$ is the plastic deformation, and $F^e$ is the elastic deformation.
We use the framework of the phase-field method for fracture~\cite{Bourdin:2008a,Karma:2001a} by
introducing the fracture phase field $\phi$ along with the process zone size $\xi$. 
To forbid interpenetration of the fracture faces similar to~\cite{Borden:2016}, we write the free energy with using an isochoric-volumetric split  in undeformed coordinates as:
\begin{align}\label{eq:free-energy}
	\F(F^e,\psi,\phi)=&\int_{\Omega_0} \left((\phi^2+\eta_\xi)W^+(F^e,\psi)+W^-(F^e,\psi)\right)\,dx\nonumber\\+&\frac{3G_c}{8}\int_{\Omega_0}\left(\frac{1-\phi}{\xi}+\xi|\nabla_0\phi|^2\right)\,dx\nonumber\\+&e_0\int_{\Omega_0}\left(f_{dw}(\psi)+w^2|\nabla_0\psi|^2\right)\,dx
\end{align}
where ${\nabla_0 \bullet}=(\partial \bullet/\partial X_i)$, $G_c$ is the fracture energy and $e_0$ is the energy cost phase change for unit undeformed volume, and $w$ is the characteristic phase change thickness. 
Furthermore, we assume a neo-Hookean hyper-elastic material
\begin{equation}\label{eq:Wp}
	W^+(F^e,\psi)
\begin{cases}
	\dfrac{\mu(\psi)}{2}\left(\hat{b}^e_{kk}-2\right)+\dfrac{\kappa(\psi)}{4}\left({J^e}^2-2\ln(J^e)-1\right)& \mathrm{if}\,J^e\geq1\\
	\dfrac{\mu(\psi)}{2}\left(\hat{b}^e_{kk}-2\right)&\mathrm{otherwise}
	\end{cases}
\end{equation}
\begin{equation}\label{eq:Wm}
	W^-(F^e,\psi)=
\begin{cases}
	0& \mathrm{if}\,J^e\geq1\\
	\dfrac{\kappa(\psi)}{4}\left({J^e}^2-2\ln(J^e)-1\right)&\mathrm{otherwise}
	\end{cases}
\end{equation}
where for $J^e=det(F^e)$, we define isochoric left Cauchy-Green deformation tensor $\hat{b}^e_{ij}={F^e_{ik}F^e_{jk}}/{J^e}$. 
In this model, the shear modulus $\mu(\psi)=\psi \mu_a+(1-\psi)\mu_c$ and bulk modulus $\kappa(\psi)=\psi \kappa_a+(1-\psi)\kappa_c$ are extrapolated between the shear and bulk moduli of amorphous ($\mu_a,\kappa_a$) and crystalline phase ($\mu_c,\kappa_c$).
Following classic $J_2$ plasticity we assume that the von~Mises equivalent stress $\tau_{eq}=\sqrt{3s_{ij}s_{ij}/2}\leq(\sy+K\alpha)$ where $s_{ij}=\mu(\psi)(\hat{b}^e_{ij}-\hat{b}^e_{kk}\delta_{ij}/N)$ is the deviatoric part of the Kirchhoff stress tensor $\tau$ ($\tau=J\sigma$ where $\sigma$ is the Cauchy stress tensor), $\sy$ is the yield strength and $\alpha$ is the isotropic hardening parameter (at the infinitesimal strain limit the isotropic hardening parameter reduces to the equivalent plastic strain).
We write the governing equations for the displacement and fracture phase field as the minimizers of \eqref{eq:free-energy}:
\begin{align}
	\frac{\delta \F}{\delta u_i}&=0\quad s.t.\,\sqrt{3s_{ij}s_{ij}/2}\leq(\sy+K\alpha)\label{eq:equillibrium-u}\\
	\frac{\delta \F}{\delta \phi}&=0\label{eq:equillibrium-phi}
\end{align}
where $\delta \F/\delta \bullet$ is the Fr\'echet derivative of the free energy $\F$ with respect to field $\bullet$.
Moreover, we model the curvature-independent interface-reaction-controlled anisotropic motion of the c-Si/a-Si interface as~\cite{Nguyen:2010}:
\begin{equation}\label{eq:phase-transformation}
	\frac{\partial\psi}{\partial t}=-M\left(\frac{\nabla_0 \psi}{|\nabla_0 \psi|}\right)\left(\frac{1}{e_0}\frac{\delta \F}{\delta \psi}-w^2|\nabla_0\psi|\nabla_0\cdot\left(\frac{\nabla_0 \psi}{|\nabla_0 \psi|}\right)\right)
\end{equation}
where $M({\nabla_0 \psi}/{|\nabla_0 \psi|})$ is the anisotropic mobility of phase change where we use the same form of the anisotropic mobility $M({\nabla_0 \psi}/{|\nabla_0 \psi|})$ as An et al.~\cite{An:2015}.

\subsection*{Vulnerable window of yield strength}

\begin{figure}[htb!]
\centering
\includegraphics[width=\columnwidth]{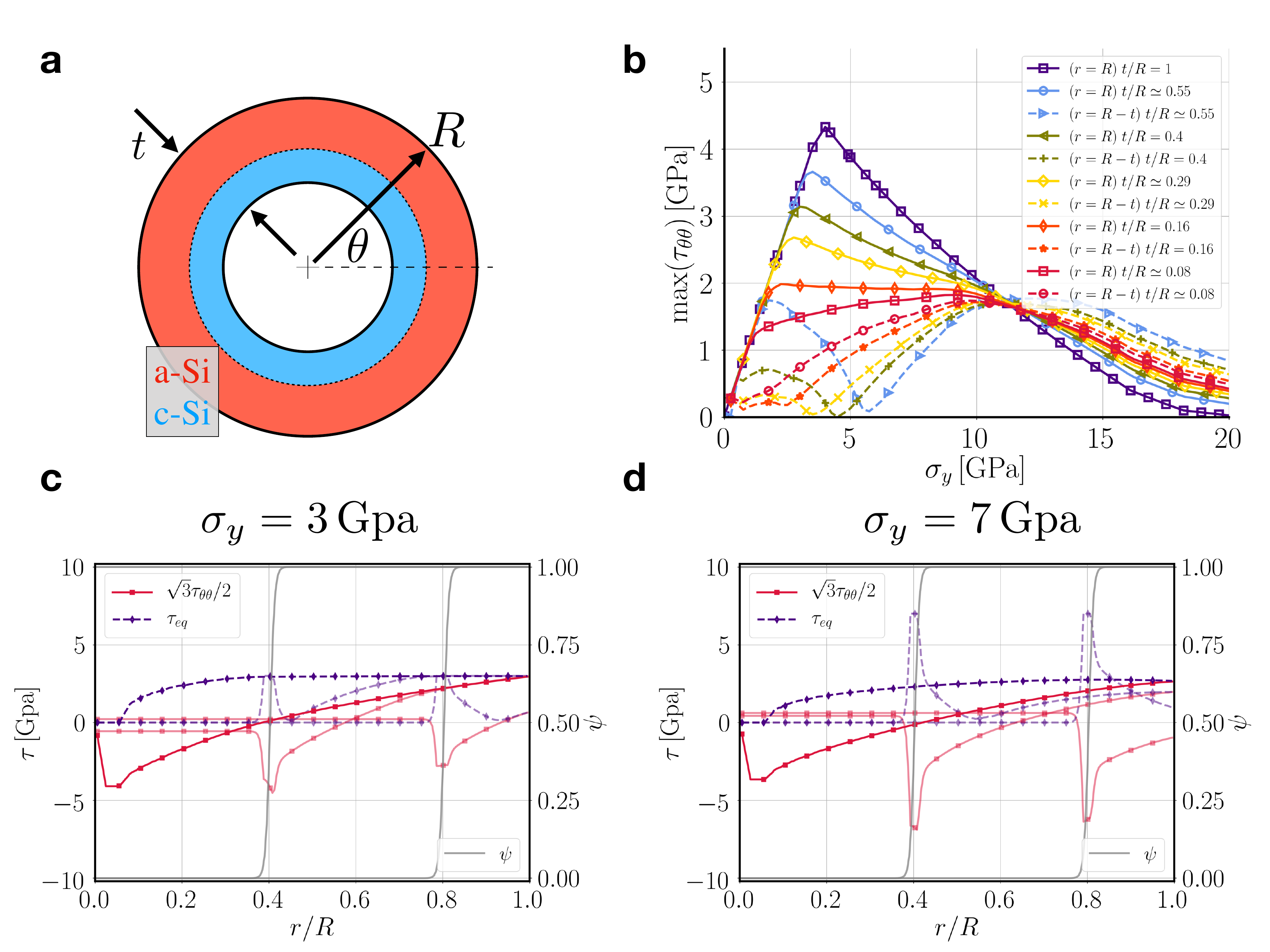}
\caption{Results of axisymmetric simulations of lithiation of hollow cylindrical crystalline nanopillars of outer radius $R=\SI{85}{\nano\meter}$ and variable thickness $t$. 
\textbf{(a)} Schematic representation of the hollow pillar geometry. 
\textbf{(b)} Plots of maximum hoop stress reached during complete lithiation on the nanopillar outer ($r=R$, solid lines) and inner ($r=R-t$, dashed lines) boundaries versus yield strength $\sy$ for different $t/R$ ratios. Plots predict the existence of a ``vulnerable window'' of $\sy$ inside which the maximum hoop stress can exceed the threshold for fracture.
\textbf{(c)} Radial profiles at different times of the phase transformation $\psi$ field (right vertical axis, gray lines with the c-Si/a-Li$_x$Si interface   located at $\psi=0.5$), the Kirchhoff hoop stress $\sqrt{3}\tau_{\thth}/2$ (left vertical axis, red lines with earlier stages shown using lighter red) and Von Mises stress $\tau_{eq}$ (left vertical axis, blue lines) for $\sy=\SI{3}{\giga\pascal}$. 
Plots show compressive yielding followed by subsequent reversal of the sign of the hoop stress and yielding under tension.
\textbf{(d)} Same as  \textbf{(c)} but for higher yield strength $\sy=\SI{7}{\giga\pascal}$ where tensile yielding does not occur.}
\label{fig:figure1}
\end{figure}

To understand the basic mechanism of stress generation in these components, as the first step, we performed an exhaustive series of axisymmetric simulations without fracture (i.e simulations in which stress fields and the plastic flow hardening parameter are assumed to only vary radially). 
Results of the axisymmetric computations are shown in Figure~\ref{fig:figure1}.
Figure~\ref{fig:figure1}(c-d) shows the evolution of the hoop stresses and equivalent von~Mises stress during lithiation of a solid nanopillar where we can identify three regimes in these figures.
As the phase transformation boundary ($\psi=0.5$) invades inside the particle from the outer boundary ($r=R$), it creates compressive stresses due to the large volumetric expansion of the a-Li$_x$Si phase.
The resulting compressive stresses generate plastic flow that caps the von~Mises stress at $\sy$.
As the crystalline core shrinks further, the compressive stresses on the outer boundary subside and change sign due to the initial compressive yield.
Consequently, 
the hoop stress on the outer boundary changes sign and becomes tensile (\Fig~\ref{fig:figure1}(c-d)), thereby confirming the knock-on effect of compressive yielding on the creation of tensile stresses that has been hypothesized to cause cracking~\cite{Liu:2011a,Yang:2012a,Yang:2014,An:2015}. 
Importantly, for $\sy$ smaller than approximately $\SI{4}{\giga\pascal}$ for the present parameters, the tensile hoop stress reaches the yield strength before the c-Si core has vanished, which results in secondary plastic yielding under tension (\Fig~\ref{fig:figure1}c). 
In this range ($\sy\le \SI{4}{\giga\pascal}$), the maximum hoop stress reached during complete lithiation, $\max(\tau_{\thth})$, increases linearly with $\sy$ as shown in \Fig~\ref{fig:figure1}b for $t/R=1$ corresponding to a solid pillar; since the outer boundary is traction free ($\tau_{rr}\equiv0$) and the von~Mises stress is capped by $\sy$, $\max(\tau_{\thth})=2\sy/\sqrt{3}$ on that boundary for plane-strain. 
In contrast, for larger $\sy$, compressive yielding requires a larger lithiated fraction, which reduces the amount of volumetric expansion available to create tensile stresses during shrinkage of the remaining c-Si core.  Therefore, $\max(\tau_{\thth})$ remains below the yield strength and decreases with increasing $\sy$ as shown in \Fig~\ref{fig:figure1}b.
We should also highlight that although all simulations presented in this article were performed using $\beta=0.7$ that corresponds to $\sim280\%$ volume change at full lithiation, our results show that there exists a universal relationship between the dimensionless maximum hoop stress  $\max(\tau_{\thth})/\mu_a \beta$ and  the dimensionelss yield strength $\sy/\mu_a \beta$ (see \Fig~\myzref{sfig:stt-vs-sy}), such that our findings can be extended to other materials whose phase-transformation result in smaller volume changes.
This universality can be readily understood noticing that at smaller expansion coefficients, smaller stresses are generated; therefore, the knock-on effect of the compressive yielding only takes place at smaller yield strength.
Crucially, these results show that the vulnerable window of yield strength is shifted to smaller values of yield strength for smaller expansion coefficient.
The resulting hat shape of the $\max(\tau_{\thth})$ versus $\sy$ plot in \Fig~\ref{fig:figure1}b suggests the existence of a vulnerable window for fracture corresponding to the range of $\sy$ where the maximum hoop stress becomes large enough to initiate fracture.
Specifically, existence of a maximum generated hoop stress at a critical yield strength increases the available energetic driving force for fracture at lower yield strength as confirmed below by our full 2D simulations (Figures~\ref{fig:figure3}--\ref{fig:figure5}).
\Fig~\ref{fig:figure1}b also shows plots of $\max(\tau_{\thth})$ versus $\sy$ on the inner ($r=R-t$) and outer ($r=R$) boundaries of hollow nanopillars. The maximum hoop stress on the outer boundary still exhibits a maximum. However, since the compliance of the annulus is inversely related to its slenderness $t/R$, the maximum stresses on both boundaries decrease with increasing slenderness. 

\begin{figure}[htb!]
\centering
\includegraphics[width=\columnwidth]{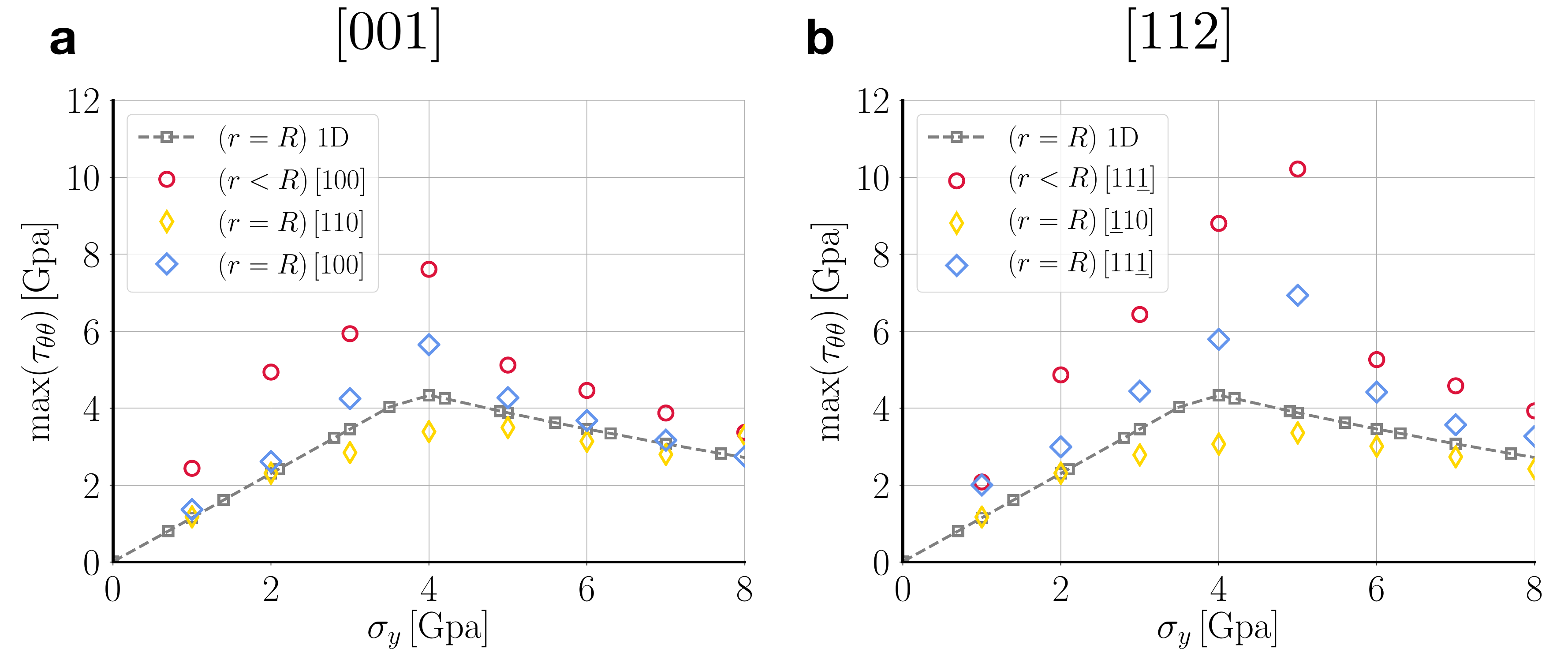}
\caption{
Comparison of 2D and 1D axisymmetric simulations of lithiation of solid nanopillars of radius $R=\SI{85}{\nano\meter}$ without fracture. Plots of maximum hoop stress $\tau_{\thth}$ vs. yield stress $\sy$ for $[001]$ \textbf{(a)} and [112] \textbf{(b)} oriented nanopillars ($r$ is the radial coordinate in the undeformed frame). 
In 2D simulations, localization of plastic flow at orientations corresponding to sharp corners of the crystalline Si core and concomitant creation of V-shaped notches (\Fig~\ref{fig:figure3}) magnifies the magnitude of stresses, thereby enlarging the size of the vulnerable window for fracture.
The largest stresses are created along the crystalline corners at a small distance from the surface (red circles).
}
\label{fig:figure2}
\end{figure}

Next, we performed 2D simulations of anisotropic swelling without fracture 
for pillars oriented in two crystallographic directions $[001]$ and $[112]$.
In these simulations, the crystalline core is no longer circular and the anisotropic mobility of the amorphization front creates a crystalline silicon core with sharp corners.
During lithiation, those crystalline corners concentrate stresses and localize plastic flow in their vicinity.
When the stresses change sign and become tensile on the pillar outer boundary, shear localization produces V-shaped notches at orientations corresponding to these corners for lower yield strength, which allows tensile yielding to occur on the periphery subsequent to compressive yielding, but not larger yield strength where tensile yielding does not occur. This difference can be seen in the pillar morphologies in \Fig~\ref{fig:figure3} for $\sy=\SI{1}{\giga\pascal}$ and $\sy=\SI{10}{\giga\pascal}$ ($\sy=\SI{7}{\giga\pascal}$ for $[112]$ oriented pillar) that did not fracture.
Those notches further concentrate stresses, thereby augmenting the magnitude of hoop stresses several fold at those orientations. This magnification is shown in \Fig~\ref{fig:figure2} where we compare $\max(\tau_{\thth})$, defined as before as the maximum hoop stress reached in time during complete lithiation, from 1D axisymmetric computations of isotropic swelling and the present 2D computations of anisotropic swelling. For the latter case, we report $\max(\tau_{\thth})$ both on the outer surface (blue and yellow diamonds) and at a position inside the particle close to the outer surface (red circles) where $\max(\tau_{\thth})$ reaches its maximum value along a vertical axis that contains the corners of the c-Si core. The maximum hoop stress is seen to be magnified both by localization of plastic deformation during compressive yielding, which occurs for all $\sy$ reported, and V-shape notches that form for lower $\sy$ due to tensile yielding. 

The 2D results in \Fig~\ref{fig:figure2} confirm the existence of a critical yield strength that generates maximal tensile stresses but were obtained from simulations without fracture. 
To investigate the effect of stress augmentation on fracture, we repeated a series of simulations with fracture for different $\sy$ and the $[001]$ and $[112]$ crystallographic orientations and for fixed process zone size to radius ratio $\xi/R=0.02$.
The time evolutions of particle morphologies in  \Fig~\ref{fig:figure3} show that nanopillar fracture during lithiation over an intermediate range of $\sy$ inside the vulnerable window centered at the critical yield strength.
For $\sy=\SI{1}{\giga\pascal}$, V-shaped notches are created along the crystalline corner directions but the magnitude of surface tensile stresses are insufficient for crack initiation.
For $\sy=\SI{3}{\giga\pascal}$, cracks initiate due to stress concentration at V-shaped notches and propagate unstably towards the crystalline core.
For $\sy=\SI{5}{\giga\pascal}$, tensile yielding and hence V-shaped notches are absent but tensile stresses grow sufficiently large (at a later stage of charging compared to $\sy=\SI{3}{\giga\pascal}$) to create two pairs of cracks that propagate unstably after initiation. 
The horizontal pair subsequently arrests and the vertical pair propagates further due to spontaneous symmetry breaking.
Finally, for $\sy=\SI{10}{\giga\pascal}$ ($\sy=\SI{7}{\giga\pascal}$ for $[112]$ oriented pillar), reduced compressive plastic flow in the system suppresses the subsequent increase of tensile stresses, thereby preventing crack initiation.
In $[001]$ oriented pillars, crack initiate at the four corners corresponding to $[100]$ and $[010]$ directions where stresses are largest. In $[112]$ oriented pillars, crack initiate at the two corners along $[11\underline{1}]$ directions.

\begin{figure}[htb!]
\centering
\includegraphics[width=\columnwidth]{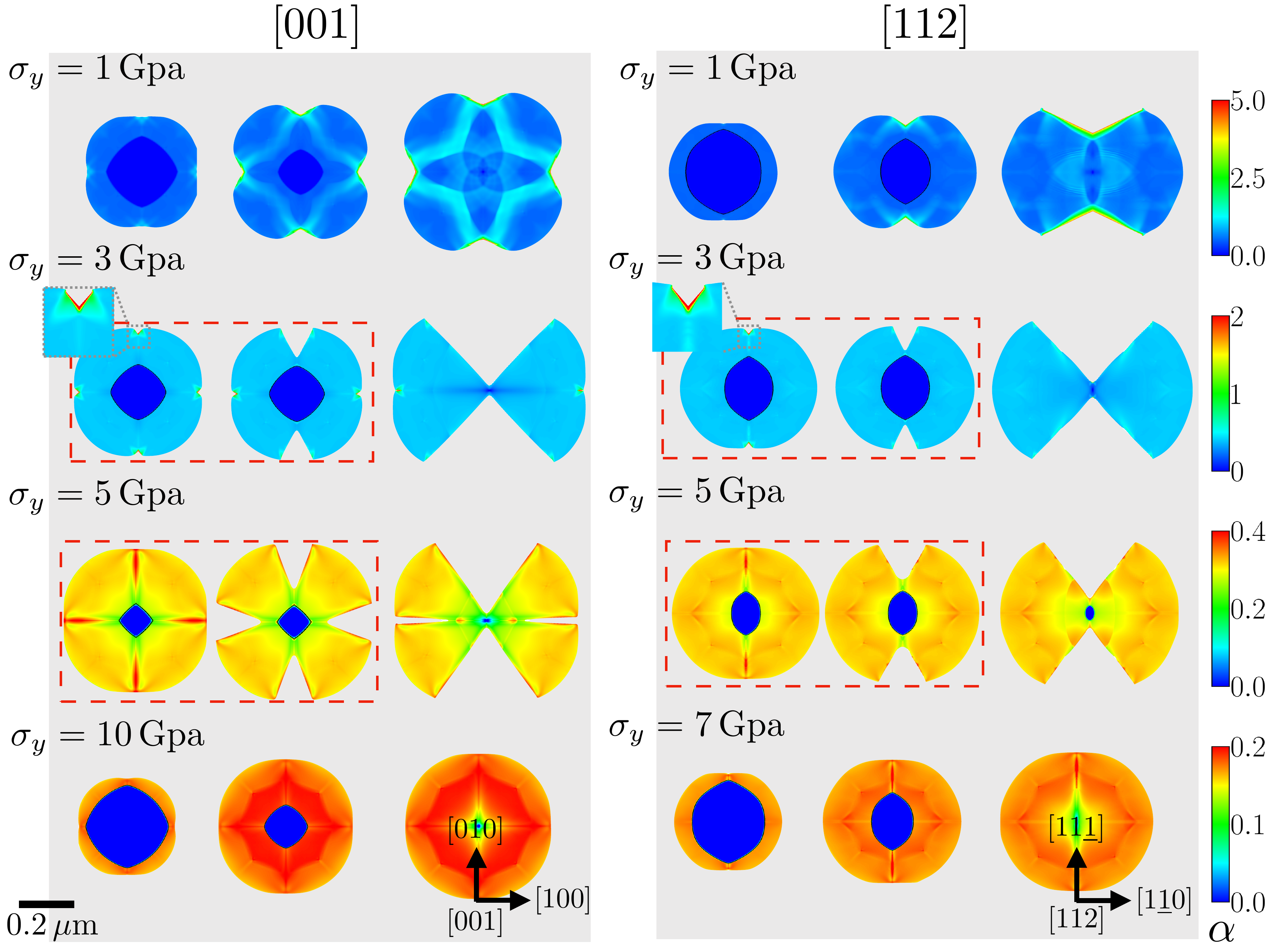}
\caption{Phase-field simulation of fracture during the lithiation of $[001]$ \textbf{(left)} and $[112]$ \textbf{(right)} oriented $R=\SI{85}{\nano\meter}$ nanopillars for different $\sy$ and for fixed process zone size to radius ratio $\xi/R=0.02$.
Color map depicts of the hardening parameter $\alpha$ and the thick black line shows the amorphous-crystalline boundary (\ie $\psi=0.5$ contour line).
The red dashed boxes show snapshots just before crack initiation and after unstable penetration towards the crystalline core. The results confirm the prediction of a window of $\sy$ for fracture and distinguishes modes of fracture with (second row) and without (third row) localization of plastic deformation creating a surface V-shaped notch prior to fracture (see also SI movies 1a,b to 6a,b).}
\label{fig:figure3}
\end{figure}

\subsection*{Size effects}

Experimental observations of Si anodic components show a clear size dependency where, for example, nanospheres with radii smaller than $\sim \SI{75}{\nano\meter}$~\cite{Liu:2012} and nanopillars  with radii smaller than $\SI{120}{\nano\meter}$~\cite{Lee:2012} do not break during lithiation.
Such size dependencies, prevalent in brittle and quasi-brittle materials~\cite{Lawn:1976,Alkadi:2016,Aliha:2017,Bazant:2019}, are typically characterized by a power law relationship between the stress to fracture and component size, $\tau_c\sim 1/\sqrt{R}$, for $R$ much larger than the process zone size.
Within the theoretical framework of Linear Elastic Fracture Mechanics (LEFM), which treats crack surfaces as sharp boundaries, this power law is readily obtained from the expression for the energy release rate at the tip of a crack of length $a$ under a spatially homogeneous critical stress $\tau_c$, which can be written as $G=\mathcal{C}a\sigma^2(1-\nu)/\mu$ for plane-strain where $\mathcal{C}$ is a dimensionless constant that generally depends on particle geometry and load configuration.  Equating the energy release rate with the fracture energy ($G=G_c$), we obtain the expression
\begin{equation}\label{eq:critical-stress}
	\tau_c=\sqrt{\frac{G_c\mu}{\mathcal{C}a(1-\nu)}}
\end{equation}
and thus the scaling $\tau_c\sim 1/\sqrt{R}$ by further assuming that the maximum flaw size $a$ increases proportionally to the particle size ($a\sim R$ and $a\ll R$). 
In the phase-field model used in this article, which describes the state of the material with a spatially varying scalar field $\phi$,
crack nucleation is an inherent property of the model and occurs via an instability that causes $\phi$ to develop a local dip ($\phi\rightarrow 0$) when the local stress exceeds a critical value $\tau_c\sim \sqrt{{G_c\mu}/\xi}$~\cite{Pham:2013} (up to a numerical prefactor that also depends on particle geometry and load configuration). 
Consequently, by comparing the above scaling expression for $\tau_c$ to Eq.~\eqref{eq:critical-stress}, we can physically interpret $\xi$ as playing an analogous role to the dominant flaw size in the LEFM framework. Furthermore, by using the result of a stability analysis of a 1D stretched strip in the phase-field model, which yields the prediction $\tau_c=\sqrt{{3G_c\mu}/{4\xi(1-\nu)}}$~\cite{Pham:2013}, we obtain the constant $\mathcal{C}=4/3$ by comparison with Eq. \ref{eq:critical-stress} with $a=\xi$. This value is close to the standard LEFM value $\mathcal{C}=\pi/2$ for a crack of length $2a$ in a uniform stress.
One main qualitative difference, however, is that crack nucleation in the phase-field model occurs through an instability of the pristine state in which $\phi$ is spatially uniform and hence does not require the introduction of a flaw in the form of a finite length seed crack as in LEFM.

\begin{figure}[htb!]
\centering
\includegraphics[width=\columnwidth]{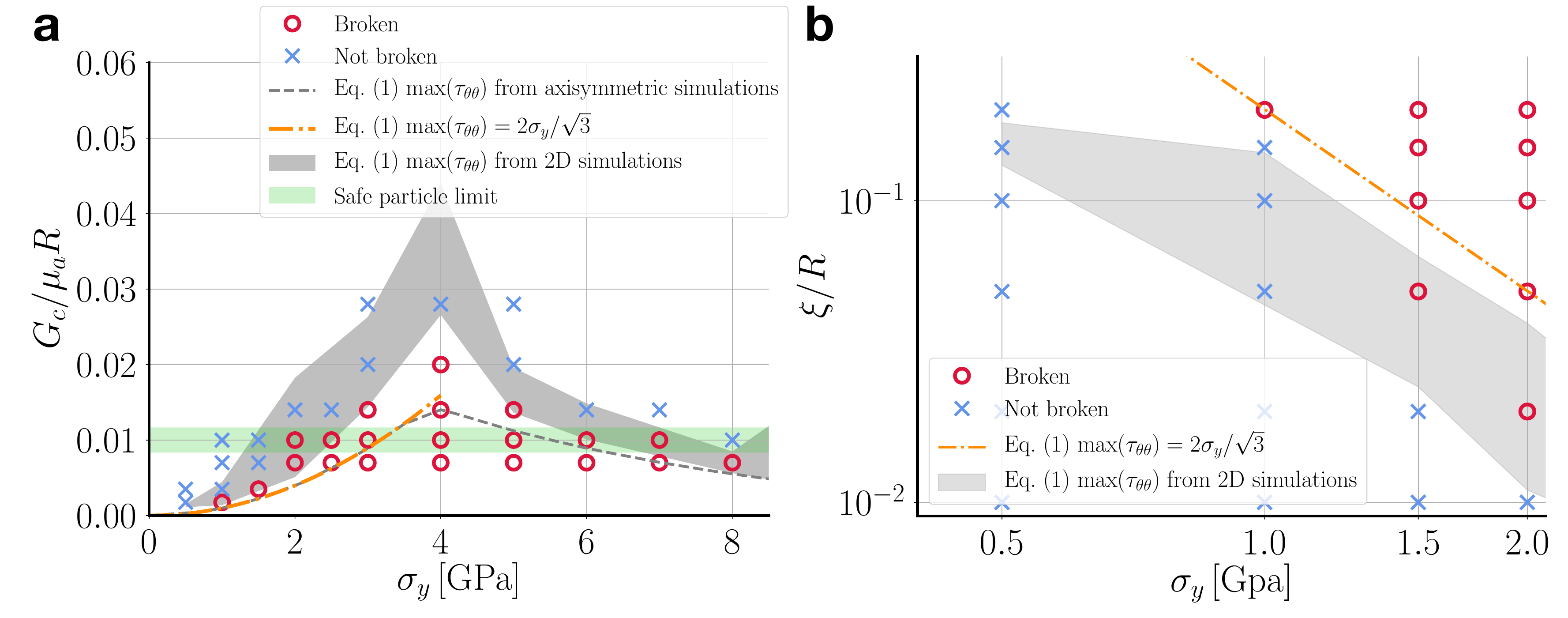}
\caption{
\textbf{(a)}
Vulnerable window of fracture plotted for yield strength $\sy$ vs. dimensionless fracture energy (particle size) $G_c/(\mu_a R)$ for $[001]$ oriented nanopillars for fixed process zone size to radius ratio $\xi/R=0.02$.
Circles depict cases with, and crosses show cases without fracture.
The results confirm the existence of a vulnerable window of fracture energy (size) and yield strength for fracture. 
Lines show the comparison with the closed-form approximation (Eq.~\eqref{eq:Gc_vs_max_hoop}) based on $\max(\tau_{\thth})$ using full 2D simulation results (red ciricles in \protect{\Fig~\ref{fig:figure2}a}, black line), 1D axisymmetric results shown in \protect{\Fig~\ref{fig:figure1}b} (dashed gray line), and $\max(\tau_{\thth})=2\sy/\sqrt{3}$ (dashed orange line). The green shaded region shows the approximate value of the dimensionless fracture energy calculated for the safe particle size $\SI{120}{\nano\meter}$~\cite{Lee:2012} and fracture energies in the range $\SIrange{5}{7}{\joule\per\meter\squared}$~\cite{Pharr:2013}.
\textbf{(b)} Phase diagram of fracture in the plane of dimensionless process zone size $0.01\le \xi/R\le 0.2$ and yield strength $\sy$ obtained
 from full 2D simulation using $G_c/(\mu_a R)=0.01$ (corresponding to $R=\SI{120}{\nano\meter}$ and fracture energy $G_c=\SI{6}{\joule\per\meter\squared}$). The results are consistent with the experimentally estimated range $\sigma_y=0.5-2$ GPa for fracture.
}
\label{fig:figure4}
\end{figure}

We take advantage of this property to investigate the particle size dependence of fracture onset by performing simulations at fixed process zone size to particle size ratio $\bar{\xi}=\xi/R$.
From the dimensional analysis, it is natural to define the dimensionless fracture energy as $G_c/(\mu_a R)$ where $\mu_a$ is the shear modulus of the amorphous a-\ce{Li_xSi} phase.
The dimensionless fracture energy can be readily interpreted as the ratio of the Griffith length scale $G_c/\mu_a$ and the nanopillar radius.
\Fig~\ref{fig:figure4}a reports the results of an extensive series of 2D phase-field fracture simulations for $[001]$ oriented nanopillars for $\bar{\xi}=0.02$, which identifies regions of the two-dimensional parameters space $G_c/(\mu_a R)$ and $\sy$ where fracture does (red circles) or does not (blue crosses) occur.
These results confirm the existence of a vulnerable window of $\sy$ where cracks initiate and propagate in nanopillars during lithiation. This window is centered around the critical value of $\sy$ generating maximal tensile stresses and shrinks in size with increasing $G_c/(\mu_a R)$.

We now assess if the onset of fracture in 2D phase-field simulations can be predicted within Griffith theory by using Eq.~\eqref{eq:critical-stress} together with values of the maximum hoop stress during lithiation obtained in 1D or 2D phase-field simulations without fracture  (\Fig~\ref{fig:figure2}).
For this, we assume stress-free surfaces ($\tau_{rr}(R)=\tau_{r\theta}(R)=0$) and small elastic strains, which allows to 
use a quadratic approximation for the stored elastic energy $W^+\simeq\tau_{\thth}^2(1-\nu_a^2)/2\El_a$ where $\El_a=9\kappa_a\mu_a/(3\kappa_a+\mu_a)$, $\nu_a=(3\kappa_a-2\mu_a)/(6\kappa_a+2\mu_a)$ are the elastic modulus and Poisson ratio of the a-Si.
We found that this small strain quadratic form provides an accurate description of the elastic energy in phase-field simulations calculated using nonlinear neo-Hookean elasticity (Eq.~\eqref{eq:Wp}).
We can then rewrite Eq.~\eqref{eq:critical-stress} as a function of the maximum hoop stress as
\begin{equation}\label{eq:Gc_vs_max_hoop}
	\frac{G_c}{\mu_a R}=\mathcal{C}{\bar{\xi}(1-\nu)}\left(\frac{\max(\tau_{\thth})}{\mu_a}\right)^2
\end{equation}
where $\mathcal{C}=4/3$. 
Substituting in the above expression the values of $\max(\tau_{\thth})$ obtained from 1D axisymmetric simulations, we obtain the gray dashed line in \Fig~\ref{fig:figure4}a that falls significantly below the boundary, comprised between red circles and blue crosses, corresponding to the onset of fracture in 2D simulations with fracture. 
Consequently, Griffith theory with axisymmetric tensile stresses underestimates the critical value of $G_c/(\mu_a R)$ for fracture at fixed $\sy$ and, hence, overestimates the safe pillar radius. 
This discrepancy can be attributed to the fact that the 1D axisymmetric simulations lack the stress amplification due to plastic localization and instability. 
We therefore conclude that localization of plasticity caused by anisotropic volumetric expansion plays a significant role in fracture of Si nanopillars. 
This conclusion is further supported by the finding that the prediction of Eq.~\eqref{eq:Gc_vs_max_hoop} is significantly improved when we use values of maximum hoop stresses obtained from 2D simulations without fracture, which exhibit stress concentration at V-shaped notches. 
Unlike in 1D axisymmetric simulations, where the tensile hoop stress is always maximum at the pillar surface, hoop stresses in 2D simulations with localization of plastic deformation reach their maximal values inside the particle at a short distance away from the V-shaped notch (\Fig~\ref{fig:figure2}). Therefore, we can reasonably use Eq.~\eqref{eq:Gc_vs_max_hoop} together with values of $\max(\tau_{\thth})$ both inside the particle and at the tip of the V-shaped notch (corresponding to the red circles and blue squares in \Fig~\ref{fig:figure2}a, respectively) to obtain lower and upper bounds for the fracture boundary in the plane of $G_c/(\mu_a R)$ and $\sy$. 
The results are depicted by the gray shaded region in  \Fig~\ref{fig:figure4}a that is comprised between the lower and upper bounds computed in this fashion. 
The fracture boundary between red circles and blue crosses in 2D phase-field simulations falls for the most part inside this gray shaded region (in particular over the range $\sy\sim \SIrange{0.5}{2}{\giga\pascal}$ of experimental relevance), thereby confirming that stress concentration near V-shaped notches is an important mechanism promoting fracture.


We can now relate our numerical findings to experimental observations of safe nanopillar sizes.
\Fig~\ref{fig:figure4}a shows that the safe nanopillar radius (where pillars with radii less than the safe value do not break) decreases with increasing $\sy$ over the estimated range $\sy=\SIrange{0.5}{2}{\giga\pascal}$ for a-\ce{Li_xSi}.
The experimental range of safe nanopillar radius is highlighted by the green shaded region in \Fig~\ref{fig:figure4}a.
This region was computed using the experimentally observed safe nanopillar radius ($\SI{120}{\nano\meter}$)~\cite{Lee:2012} and the estimated range of fracture energy ($\SIrange{5}{7}{\joule\per\meter\squared}$) from experimental measurements~\cite{Pharr:2013}. 
2D phase-field simulations predict that, inside this green shaded region of \Fig~\ref{fig:figure4}a, fracture occurs for $\sy$ between $\SI{1.5}{\giga\pascal}$  (blue crosses) and $\SI{2}{\giga\pascal}$ (red circles), which falls in the upper part of the range $\sy=\SIrange{0.5}{2}{\giga\pascal}$ estimated from experimental measurements~\cite{Sethuraman:2010,Chon:2011,Pharr:2013}. Phase-field modeling prediction also depend generally on flaw size through the ratio $\xi/R$.
To test this dependence, we repeated a series of simulations by varying $\xi/R$ over the range $0.01$ to $0.2$, which encompasses the value $0.02$ used in all simulations presented so far.
These simulations were carried out at fixed dimensionless fracture energy $G_c/(\mu_a R)=0.01$ calculated using the average reported fracture energy for a-\ce{Li_xSi} $G_c=\SI{6}{\joule\per\meter\squared}$~\cite{Pharr:2013} and the observed safe nanopillar radius $R\simeq\SI{120}{\nano\meter}$~\cite{Lee:2012}. 
The results reported in \Fig~\ref{fig:figure4}b show that nanopillars become more vulnerable to fracture with increasing $\xi/R$ as theoretically expected. For the largest value $\xi/R=0.2$ studied here, fracture occurs for $\sy$ between $\SI{0.5}{\giga\pascal}$  (blue cross) and $\SI{1}{\giga\pascal}$ (red circle), which falls in the lower part of the range $\sy=\SIrange{0.5}{2}{\giga\pascal}$ estimated from experimental measurements~\cite{Sethuraman:2010,Chon:2011,Pharr:2013}. 
While the precise value of $\xi/R$ is not known, its value is presumably much smaller than unity given that nanopillars do not typically exhibit large visible flaw sizes prior to lithiation~\cite{Liu:2012a,Lee:2012}.

We can further confirm the above estimate of $\sy$ by redoing the calculations for isotropic lithiation of amorphous Si.
Experimental observations demonstrate that a-Si lithiates isotropically~\cite{McDowell:2013,Wang:2013,Berla:2014}. 
It is therefore reasonable to use our 1D axisymmetric simulations to estimate the magnitude of hoop stresses generated during lithiation of a-Si. We can simplify Eq.~\eqref{eq:Gc_vs_max_hoop} further using the expression $\max(\tau_{\thth})\simeq 2\sy/\sqrt{3}$ valid for small $\sy$ (see \Fig~\ref{fig:figure1}b), which yields the prediction ${G_c}/{\mu_a R}=(4\mathcal{C}/3){\bar{\xi}(1-\nu)}({\sy}/{\mu_a})^2$ for isotropic lithiation of a-\ce{Li_xSi}.
Zhao~et~al.~\cite{Zhao:2012} obtained a similar expression of the form of ${G_c}/{\mu_a R}\sim({\sy}/{\mu_a})^2$ previously by an analysis of lithiation that only considers plastic deformation and computes the dimensionless prefactor numerically assuming an initial flaw size comparable to $R$. 
In contrast, here, the prefactor is obtained analytically from the aforementioned 1D stability analysis of crack initiation in the phase-field model~\cite{Pham:2013}.  
We can use this isotropic estimate along with the safe experimentally observed safe nanopillar radius $\SI{1}{\micro\meter}$~\cite{Berla:2014} to calculate an upper bound for its yield strength $\SIrange{0.4}{1.2}{\giga\pascal}$ using the process zone size $0.02\leq\xi/R\leq0.2$.
We can use a similar analysis for Ge. 
Since lithiation of Ge is observed to be isotropic, using the estimates of its shear modulus $\mu_a\simeq\SI{19}{\giga\pascal}$ we can estimate $\sy$ in the range $\SIrange{1.5}{4.6}{\giga\pascal}$.
We calculated the above range similarly using the process zone size $0.02\leq\xi/R\leq0.2$ and based on observed safe nanopillar radius of $\SI{250}{\nano\meter}$~\cite{Lee:2015c}.


\begin{figure}[htb!]
\centering
\includegraphics[width=\columnwidth]{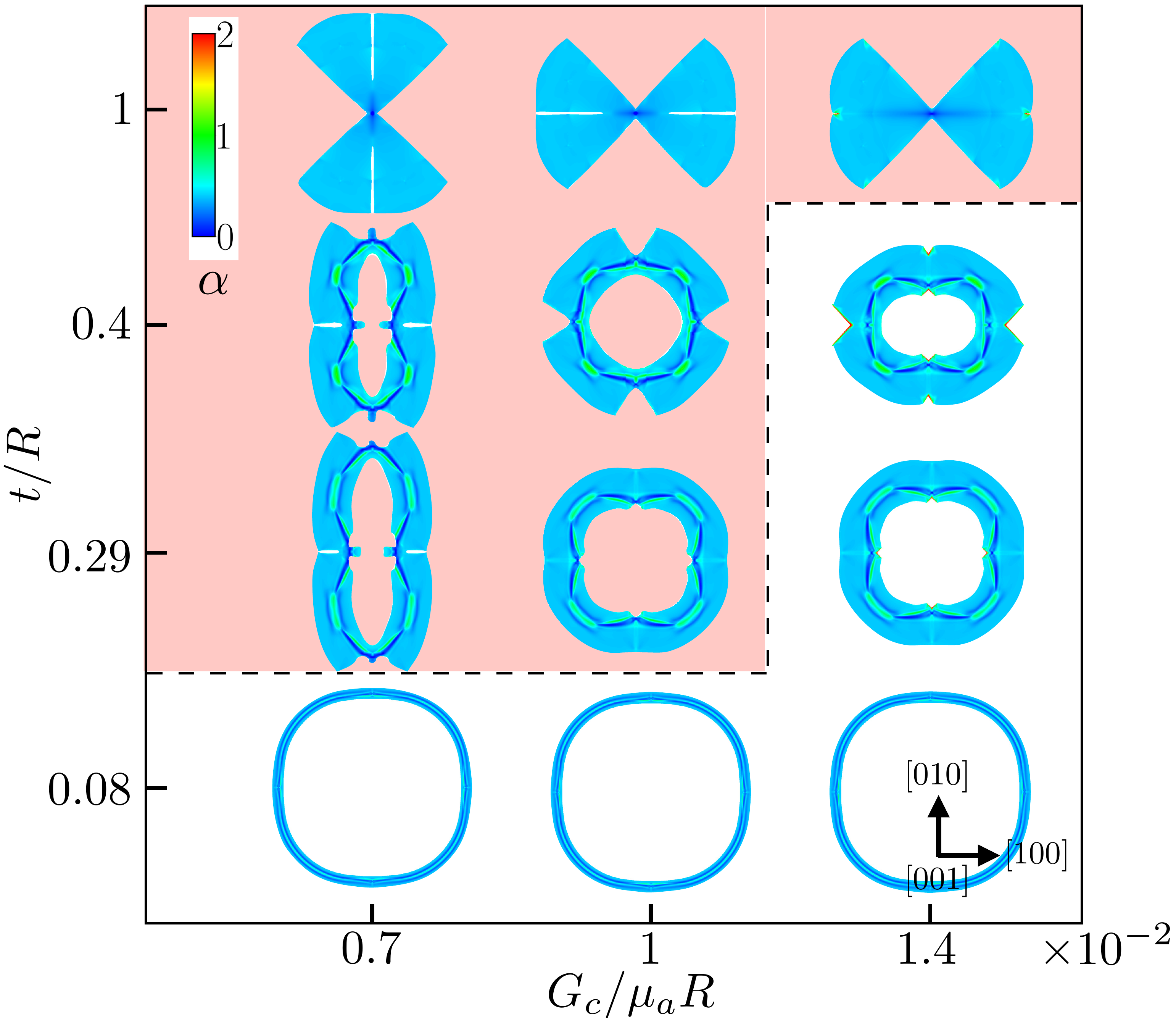}
\caption{Final cross section and fracture pattern as a function of dimensionless fracture energy $G_c/(\mu_a R)$ and slenderness $t/R$ for hollow nanopillars with equivalent cross-section area as solid nanopillars of radius $R=85, 121$ and $170$ ($G_c=\SI{6}{\joule\per\meter\squared}$, $\sy=\SI{3}{\giga\pascal}$, and $\xi/R=0.02$). Color map depicts the hardening parameter $\alpha$.
The parameter range where the nanopillar fractured is shown with the red background.
One can see that at this yield strength the more slender nanopillar mitigates the failure.
Also increasing particle size (decreasing the fracture energy) promotes initiation of cracks in-line with the experimental observations (see SI movies 7a,b and 8a,b for results of simulations for $G_c/(\mu_a R)=0.01$ and $t/R\simeq0.29$ and $t/R=0.4$).
}
\label{fig:figure5}
\end{figure}

\subsection*{Geometrical effects}

Finally, to highlight the non-trivial role of geometry beyond size effects, we investigate the fracture of hollow nanopillars that have been shown experimentally to be more resistant to fracture~\cite{Zhang:2017}.
Our axisymmetric computations predict that this geometrical protective effect is present for large enough yield strength due to a decrease of the maximum hoop stress reached during complete lithiation of crystalline Si  as a function of $\sy$ (\ie the decrease of the peak value of plots in \Fig~\ref{fig:figure1}b with increasing annulus slenderness).
However, for low yield strength, the maximum hoop stress remains bounded by $\sy$ even for large slenderness. This implies that the hollow nanopillar design can mitigate fracture only for materials with  moderately high yield strength.
To test these predictions, we modeled the lithiation and fracture of hollow nanopillars with a constant cross-sectional area equal to solid nanopillars with radii $\SIrange{85}{170}{\nano\meter}$ and different slenderness $0.08\leq t/R \leq 1$ for $\sy=\SI{3}{\giga\pascal}$.
The results illustrated in \Fig~\ref{fig:figure5} show that increased slenderness has a protective effect for this $\sy$ value. 
For example, nanopillars break for $t/R=0.4$ but only exhibit minor cracking near the interior surface of $t/R\simeq0.29$. Even more slender nanopillars ($t/R\simeq 0.08$) do not fracture. 
This protective geometrical effect, however, does not persist at lower yield strength for $\xi/R=0.02$. 
Our axisymmetric computations predict that the maximum hoop stress generated in even the thinnest annulus is equal to that of a solid nanopillar for $\sy=\SI{1}{\giga\pascal}$ (\Fig~\ref{fig:figure1}b).
Our 2D simulations confirm this prediction by showing that the hollow nanopillar for $t/R\simeq0.29$, which is protected for $\sy=\SI{3}{\giga\pascal}$, fractures for $\sy=\SI{1}{\giga\pascal}$. 

In summary, we have used a multi-physics phase-field approach to model simultaneously anisotropic phase transformation, elasto-plastic deformation, and crack initiation and propagation during lithiation of Si nanopillars. 
Our results identify a vulnerable window of yield strength inside which pillars fracture during lithiation and distinguish two different modes of fracture inside that window with and without surface localization of plastic deformation prior to fracture for lower and higher yield strength, respectively. Those two modes follow from the existence of a critical yield strength that generates maximal tensile stresses during lithiation.
Combined with experimental measurements of fracture energy~\cite{Pharr:2013} and observations of size dependent fracture~\cite{Lee:2012,Liu:2012,Berla:2014}, our results yield an estimate of yield strength within a range $\sy\simeq\SIrange{0.5}{2}{\giga\pascal}$ consistent with experimental~\cite{Sethuraman:2010,Chon:2011,Pharr:2013} and theoretical~\cite{Zhao:2011c} estimates. 
This range is smaller than the critical yield strength consistent with the observation of localized plastic deformation during lithiation of Si nanopillars~\cite{Liu:2011a,Lee:2012,Liu:2012}. 
Over this range, plastic deformation mitigates fracture by energy dissipation but, at the same time, promotes it by the creation of stress-concentrating V-shaped notches that precede quasi-brittle fracture.
Our results also suggest that the observed increased robustness of hollow Si nanopillars~\cite{Zhang:2017} is due to a reduction of the critical yield strength generating maximal tensile stresses with increasing slenderness. 
This interpretation, however, warrants further investigation since this protective effect is only significant in simulations with large enough $\sy$ values. 
The present study highlights the importance of computationally informed geometric design that takes into account the subtle interplay between material properties and geometry to generate reliable predictions of mechanical stability of high-capacity battery materials, paving the way for designs that exploit more complex geometries such as open nanoporous structures with ultra-high interfacial area~\cite{Wada:2014}.

\section*{Methods} 


All the equations are solved using Galerkin Finite Element Method. 
Furthermore, to ensure the robustness of solution only 1/4 of each geometry was simulated and appropriate boundary conditions are applied on the symmetry axises.
Our implementation is based on PETSc~\cite{Balay:2017} as the linear algebra backbone and libMesh~\cite{Kirk:2006a} for finite elements bookkeeping. 
Eq.~\eqref{eq:equillibrium-u} is solved using a Newton method where we calculate the consistent tangent moduli explicitly at each iteration.
Furthermore, Eq.~\eqref{eq:phase-transformation} is integrated explicitly using 50 substeps during each time step to ensure the accuracy and stability of integration. Table~\ref{tab:matprop} summarizes the values of different parameters used in our simulations.

\begin{table}[htb!]
\centering
\caption{Material properties of a-\ce{Si} and c-\ce{Si} used in the simulations.}
\vspace{.5em}
\begin{tabular}{l|c c}
\toprule
Material property & Value & Units \tabularnewline
\midrule
$\kappa_{c}$~\cite{An:2015} & 108 	& \si{\giga\pascal} 			\tabularnewline
$\kappa_{a}$~\cite{An:2015} & 10.8 	& \si{\giga\pascal} 			\tabularnewline
$\mu_c$~\cite{An:2015} 		& 50	& \si{\giga\pascal} 			\tabularnewline
$\mu_a$~\cite{An:2015} 		& 5		& \si{\giga\pascal} 			\tabularnewline
$K$ 						& 50	& \si{\mega\pascal} 			\tabularnewline
$G_c$~\cite{Pharr:2013}		& 6		& \si{\joule\per\meter\squared} \tabularnewline
$w$ 						& 0.85	& \si{\nano\meter} 				\tabularnewline
$\beta$ 					& 0.7	& -								\tabularnewline
\bottomrule
\end{tabular}
\label{tab:matprop}
\end{table}

\section*{Author contributions} 

\noindent A.M. and A.K. conceived the theoretical study and jointly interpreted the numerical results and wrote the paper. A.M. carried out the numerical study. 



\section*{Acknowledgements}
A.M. and A.K. acknowledge the support of Grant No. DE-FG02-07ER46400 from the U.S. Department of Energy, Office of Basic Energy Sciences.
The majority of the numerical simulations were performed on the Northeastern University Discovery cluster at the Massachusetts Green High Performance Computing Center (MGHPCC).

\bibliographystyle{unsrt}
\bibliography{Mesgarnejad-Karma-Li-Si}

\end{document}


\date{}
\maketitle


\begin{figure}[htb!]
\centering
\includegraphics[width=\textwidth]{figures/hoop_stress_vs_sy_beta.pdf}
\caption{
Universal curve showing the numerical results of 1D axisymmetric simulations relating the dimensionless maximum hoop stress $\max(\tau_{\theta\theta})/\mu_a\beta$  and different dimensionless yield stresses $\sigma_y/\mu_a\beta$ for different expansion coefficient $\beta$ in axisymmetric equations. 
}
\label{sfig:stt-vs-sy}
\end{figure}


\begin{longtable}{cc cc}
\caption{List of supplementary movies.}
\label{tab:movies}\\
\toprule
Name & Description & Figure\\
\midrule
\endhead
    \midrule
    {\footnotesize\itshape Continue on the next page}
\endfoot
    \bottomrule
\endlastfoot
\texttt{movie1a.mov} & \makecell{Evolution of hardening parameter $\alpha$ for \\$R=\SI{85}{\nano\meter}$ $[001]$ oriented nanopillar\\ at $\sigma_y=\SI{1}{\giga\pascal}$}  & \myzref{fig:figure3} \\ \addlinespace[5pt]
\texttt{movie1b.mov} & \makecell{Evolution of Kirchhoff hoop stress $\tau_{\theta\theta}\,[\mathrm{Gpa}]$ for \\$R=\SI{85}{\nano\meter}$ $[001]$ oriented nanopillar\\ at $\sigma_y=\SI{1}{\giga\pascal}$}  & \myzref{fig:figure3}  \\ \addlinespace[5pt]
\texttt{movie2a.mov} & \makecell{Evolution of hardening parameter $\alpha$ for \\$R=\SI{85}{\nano\meter}$ $[001]$ oriented nanopillar\\ at $\sigma_y=\SI{3}{\giga\pascal}$}   & \myzref{fig:figure3} \\ \addlinespace[5pt]
\texttt{movie2b.mov} & \makecell{Evolution of Kirchhoff hoop stress $\tau_{\theta\theta}\,[\mathrm{Gpa}]$ for \\$R=\SI{85}{\nano\meter}$ $[001]$ oriented nanopillar\\ at $\sigma_y=\SI{3}{\giga\pascal}$}   & \myzref{fig:figure3} \\ \addlinespace[5pt]
\texttt{movie3a.mov} & \makecell{Evolution of hardening parameter $\alpha$ for \\$R=\SI{85}{\nano\meter}$ $[001]$ oriented nanopillar\\ at $\sigma_y=\SI{5}{\giga\pascal}$}   & \myzref{fig:figure3} \\ \addlinespace[5pt]
\texttt{movie3b.mov} & \makecell{Evolution of Kirchhoff hoop stress $\tau_{\theta\theta}\,[\mathrm{Gpa}]$ for \\$R=\SI{85}{\nano\meter}$ $[001]$ oriented nanopillar\\ at $\sigma_y=\SI{5}{\giga\pascal}$}   & \myzref{fig:figure3} \\ \addlinespace[5pt]
\texttt{movie4a.mov} & \makecell{Evolution of hardening parameter $\alpha$ for \\$R=\SI{85}{\nano\meter}$ $[001]$ oriented nanopillar\\ at $\sigma_y=\SI{10}{\giga\pascal}$}   & \myzref{fig:figure3} \\ \addlinespace[5pt]
\texttt{movie4b.mov} & \makecell{Evolution of Kirchhoff hoop stress $\tau_{\theta\theta}\,[\mathrm{Gpa}]$ for \\$R=\SI{85}{\nano\meter}$ $[001]$ oriented nanopillar\\ at $\sigma_y=\SI{10}{\giga\pascal}$}  & \myzref{fig:figure3}   \\ \addlinespace[5pt]
\texttt{movie5a.mov} & \makecell{Evolution of hardening parameter $\alpha$ for \\$R=\SI{85}{\nano\meter}$ $[112]$ oriented nanopillar\\ at $\sigma_y=\SI{1}{\giga\pascal}$}   & \myzref{fig:figure3} \\ \addlinespace[5pt]
\texttt{movie5b.mov} & \makecell{Evolution of Kirchhoff hoop stress $\tau_{\theta\theta}\,[\mathrm{Gpa}]$ for \\$R=\SI{85}{\nano\meter}$ $[112]$ oriented nanopillar\\ at $\sigma_y=\SI{1}{\giga\pascal}$}   & \myzref{fig:figure3} \\ \addlinespace[5pt]
\texttt{movie6a.mov} & \makecell{Evolution of hardening parameter $\alpha$ for \\$R=\SI{85}{\nano\meter}$ $[112]$ oriented nanopillar\\ at $\sigma_y=\SI{3}{\giga\pascal}$}   & \myzref{fig:figure3} \\ \addlinespace[5pt]
\texttt{movie6b.mov} & \makecell{Evolution of Kirchhoff hoop stress $\tau_{\theta\theta}\,[\mathrm{Gpa}]$ for \\$R=\SI{85}{\nano\meter}$ $[112]$ oriented nanopillar\\ at $\sigma_y=\SI{3}{\giga\pascal}$}   & \myzref{fig:figure3} \\ \addlinespace[5pt]
\texttt{movie7a.mov} & \makecell{Evolution of hardening parameter $\alpha$ for \\$R=\SI{85}{\nano\meter}$ $[001]$ oriented hollow $t/R=0.4$ nanopillar\\ at $\sigma_y=\SI{5}{\giga\pascal}$}   & \myzref{fig:figure5}  \\ \addlinespace[5pt]
\texttt{movie7b.mov} & \makecell{Evolution of Kirchhoff hoop stress $\tau_{\theta\theta}\,[\mathrm{Gpa}]$ for \\$R=\SI{85}{\nano\meter}$ $[001]$ oriented hollow $t/R=0.4$ nanopillar\\ at $\sigma_y=\SI{5}{\giga\pascal}$}   & \myzref{fig:figure5}  \\ \addlinespace[5pt]
\texttt{movie8a.mov} & \makecell{Evolution of hardening parameter $\alpha$ for \\$R=\SI{85}{\nano\meter}$ $[001]$ oriented hollow $t/R\simeq0.29$ nanopillar\\ at $\sigma_y=\SI{10}{\giga\pascal}$}   & \myzref{fig:figure5}  \\ \addlinespace[5pt]
\texttt{movie8b.mov} & \makecell{Evolution of Kirchhoff hoop stress $\tau_{\theta\theta}\,[\mathrm{Gpa}]$ for \\$R=\SI{85}{\nano\meter}$ $[001]$ oriented hollow $t/R\simeq0.29$ nanopillar\\ at $\sigma_y=\SI{10}{\giga\pascal}$}   & \myzref{fig:figure5}   \\ \addlinespace[5pt]
\bottomrule
\end{longtable}